# Carrier–phonon decoupling via annealing enhances thermoelectric performance of $Bi_2(Te,Se)_3$  FREE

Xinxiu Cheng; Liqing Xu; Zhibin Gao ✉ ; Wei Liu; Zhanxiang Yin; Xiangdong Ding ; Yu Xiao ✉

# Carrier–phonon decoupling via annealing enhances thermoelectric performance of Bi$_2$(Te,Se)$_3$

Xinxiu Cheng,[1] Liqing Xu,[2,3] Zhibin Gao,[1,a)] 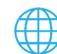 Wei Liu,[1] Zhanxiang Yin,[2] Xiangdong Ding,[4] 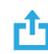 and Yu Xiao[2,a)]

### AFFILIATIONS

[1]State Key Laboratory of Porous Metal Materials, School of Materials Science and Engineering, Xi'an Jiaotong University, Xi'an 710049, China
[2]School of Materials and Energy, University of Electronic Science and Technology of China, Chengdu 611731, China
[3]Shenzhen Institute for Advanced Study, University of Electronic Science and Technology of China, Shenzhen 518110, China
[4]State Key Laboratory for Mechanical Behavior of Materials, Xi'an Jiaotong University, Xi'an 710049, China

[a)]Authors to whom correspondence should be addressed: zhibin.gao@xjtu.edu.cn and xiaoyu@uestc.edu.cn

### ABSTRACT

Thermoelectric cooling based on the Peltier effect requires high-performance materials near room temperature. While Bi$_2$Te$_3$ remains the most viable commercial thermoelectric material, n-type Bi$_2$Te$_{3-x}$Se$_x$ (BTS) lags behind its p-type counterpart due to intrinsic anisotropy. In this work, Bi$_2$Te$_{2.6}$Se$_{0.4}$ synthesized by melting-hot pressing followed by 100 h annealing (MT-HP-AN) at 723 K exhibits markedly improved performance. Annealing introduces cation vacancies via Te(Se) volatilization, lowering carrier density and enhancing mobility, while simultaneously increasing phonon scattering. A peak $ZT$ of 1.06 at 373 K and an average $ZT$ of 0.99 at 300–423 K are achieved. This MT-HP-AN approach offers a simple yet effective strategy to decouple carrier and phonon transport, advancing the potential of n-type BTS for thermoelectric cooling applications.

Published under an exclusive license by AIP Publishing. https://doi.org/10.1063/5.0293474

Thermoelectric (TE) materials, capable of directly converting heat into electricity and vice versa, are critical for solid-state cooling in emerging fields such as 5G communication and microelectronics.[1–3] The cooling efficiency of thermoelectric devices depends on the dimensionless figure of merit ($ZT$) of thermoelectric materials. $ZT$ value can be expressed as $ZT = (\sigma S^2 T)/(\kappa_{ele} + \kappa_{lat})$, where $\sigma$, $S$, $T$, $\kappa_{ele}$, and $\kappa_{lat}$ are the electrical conductivity, Seebeck coefficient, absolute temperature in Kelvin, electronic thermal conductivity, and lattice thermal conductivity, respectively.[4] In general, $S^2\sigma$ is referred to as power factor ($PF$), and the sum of $\kappa_{ele}$ and $\kappa_{lat}$ is referred to as total thermal conductivity ($\kappa_{tot}$). Therefore, achieving high $ZT$ value requires a high $PF$ value and a low $\kappa_{tot}$. However, these parameters are interdependent due to strong electron–phonon coupling. Decoupling carrier and phonon transport is therefore essential for improving thermoelectric performance.[5]

Bi$_2$Te$_3$-based compounds are known as the best thermoelectric materials near room temperature.[6] It includes both p-type Bi$_{2-x}$Sb$_x$Te$_3$ (BST) and n-type Bi$_2$Te$_{3-x}$Se$_x$ (BTS), making it an excellent candidate for large-scale commercial power generation and cooling applications.[3,7] In recent years, researchers have improved the $ZT$ value of 1.2–1.9 near room temperature in p-type BST by introducing nanostructuring[8] and all-scale dislocation.[9] However, n-type BTS lags behind its p-type counterpart, necessitating improvements for higher efficiency conversion and broader applications.[10] The reasons for the limited improvement in thermoelectric properties of n-type BTS are as follows. On the one hand, in terms of crystal structure, n-type BTS exhibits stronger crystallographic anisotropy than its p-type counterpart, leading to its electrical properties being highly sensitive to grain orientation. It has been reported that the conductivity anisotropy ratio of n-type BTS single crystals ranges from 4 to 7, which is significantly higher than the p-type counterpart (∼3).[2] The distinct anisotropy ratios reveal that the mobility of n-type alloys is more sensitive to the crystal texture. This pronounced anisotropy indicates that the carrier mobility in n-type alloys is more sensitive to crystal texture. On the other hand, n-type BTS typically contains a higher concentration of intrinsic point defects, such as vacancies and antisite defects, which have a more pronounced impact on its thermoelectric performance.[11] Therefore, the same methods that are used to optimized thermoelectric properties of p-type BST are invalid for n-type BTS.

The $ZT$ value of n-type BTS is generally enhanced by manipulating the synthesis process, such as extrinsic chemical doping and microstructure engineering.[12] However, while these approaches often suppress the lattice thermal conductivity through the introduction of multiple defects, they also tend to degrade electrical transport properties due to intensified carrier scattering.[13] More recently, the texture process has emerged as a promising strategy to improve thermoelectric performance of n-type BTS by enhancing grain orientation. Techniques such as hot extrusion, hot forging, liquid-phase hot deformation techniques, and rotary swaging method have demonstrated effectiveness in promoting grain rearrangement, thereby enhancing grain orientation and thermoelectric properties.[14] Unfortunately, these methods face considerable challenges, including the need for precise control of temperature and pressure, multiple hot extrusion deformations cycles, and limited process repeatability, which collectively hinder scalability for industrial applications. In addition, the intrinsic anisotropy of n-type BTS further complicates performance optimization. Therefore, a straightforward strategy that decouples phonon and carrier transport—while accounting for its anisotropic nature—could provide a more practical pathway for enhancing the thermoelectric performance of n-type BTS.

In this work, we demonstrate that a simple annealing strategy enables effective decoupling of carrier and phonon transport in Se-alloyed $Bi_2Te_3$, thereby significantly improving its thermoelectric properties. The n-type BTS were synthesized via a convenient and controlled melting-hot pressing (MT-HP) route. The annealing process was found to tune the carrier density and mobility of n-type BTS via the formation of cation vacancies, thereby enhancing the power factor (PF). Meanwhile, Te(Se) volatilization during annealing introduced atomic-scale defects, which strengthened phonon scattering and substantially reduced $\kappa_{lat}$. As a result, a synergistic optimization of carrier and phonon transport was achieved through annealing treatment. The $Bi_2Te_{2.6}Se_{0.4}$ sample annealed for 100 h exhibited a peak $ZT$ value of 1.06 at 373 K and an average $ZT$ of 0.99 in the range of 300–423 K.

We compared the thermoelectric properties of BTS polycrystalline samples prepared by melting combined with either spark plasma sintering (MT-SPS) or hot pressing (MT-HP), as presented in Fig. 1 and summarized in Table S1. Both methods offer advantages such as simplified operation, reduced processing time, homogeneous elemental

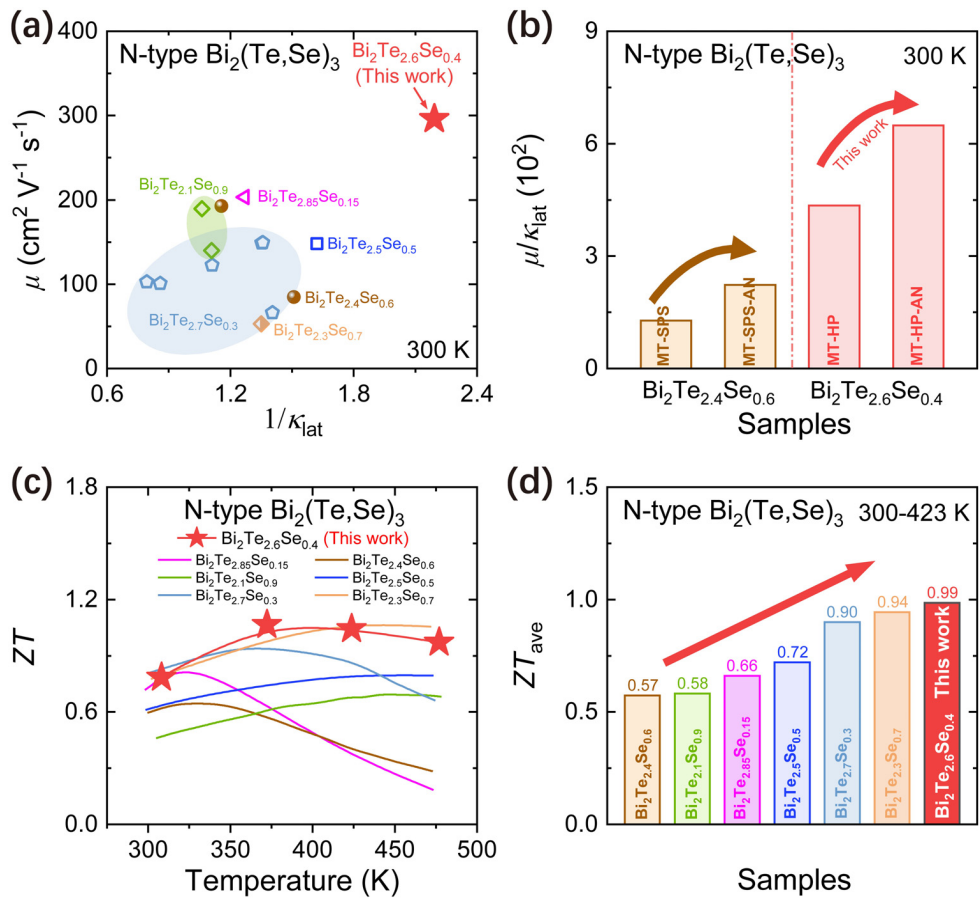

FIG. 1. Comparisons of thermoelectric transport properties in n-type $Bi_2(Te,Se)_3$-based thermoelectric materials. (a) Carrier mobility as a function of $1/\kappa_{lat}$ at room temperature (the unit of $1/\kappa_{lat}$ is $W^{-1}$ m K). Data from Refs. 12 and 14–23. (b) $\mu/\kappa_{lat}$ values of n-type $Bi_2(Te,Se)_3$ materials prepared by melting process combined with spark plasma sintering (MT-SPS) or hot pressing sintering (MT-HP) with and without annealing treatment (MT-SPS-AN or MT-HP-AN) (the unit of $\mu/\kappa_{lat}$ is $cm^2$ $V^{-1}$ $s^{-1}$/W $m^{-1}$ $K^{-1}$). Data from Refs. 20 and 21. (c) and (d) Comparisons of the temperature-dependent $ZT$ value and $ZT_{ave}$ value at 300–423 K with those of other n-type $Bi_2(Te,Se)_3$ materials, including $Bi_2Te_{2.85}Se_{0.15}$, $Bi_2Te_{2.7}Se_{0.3}$, $Bi_2Te_{2.5}Se_{0.5}$, $Bi_2Te_{2.4}Se_{0.6}$, $Bi_2Te_{2.3}Se_{0.7}$, and $Bi_2Te_{2.1}Se_{0.9}$. Data from Refs. 14, 15, 17, 19, 21, and 23.

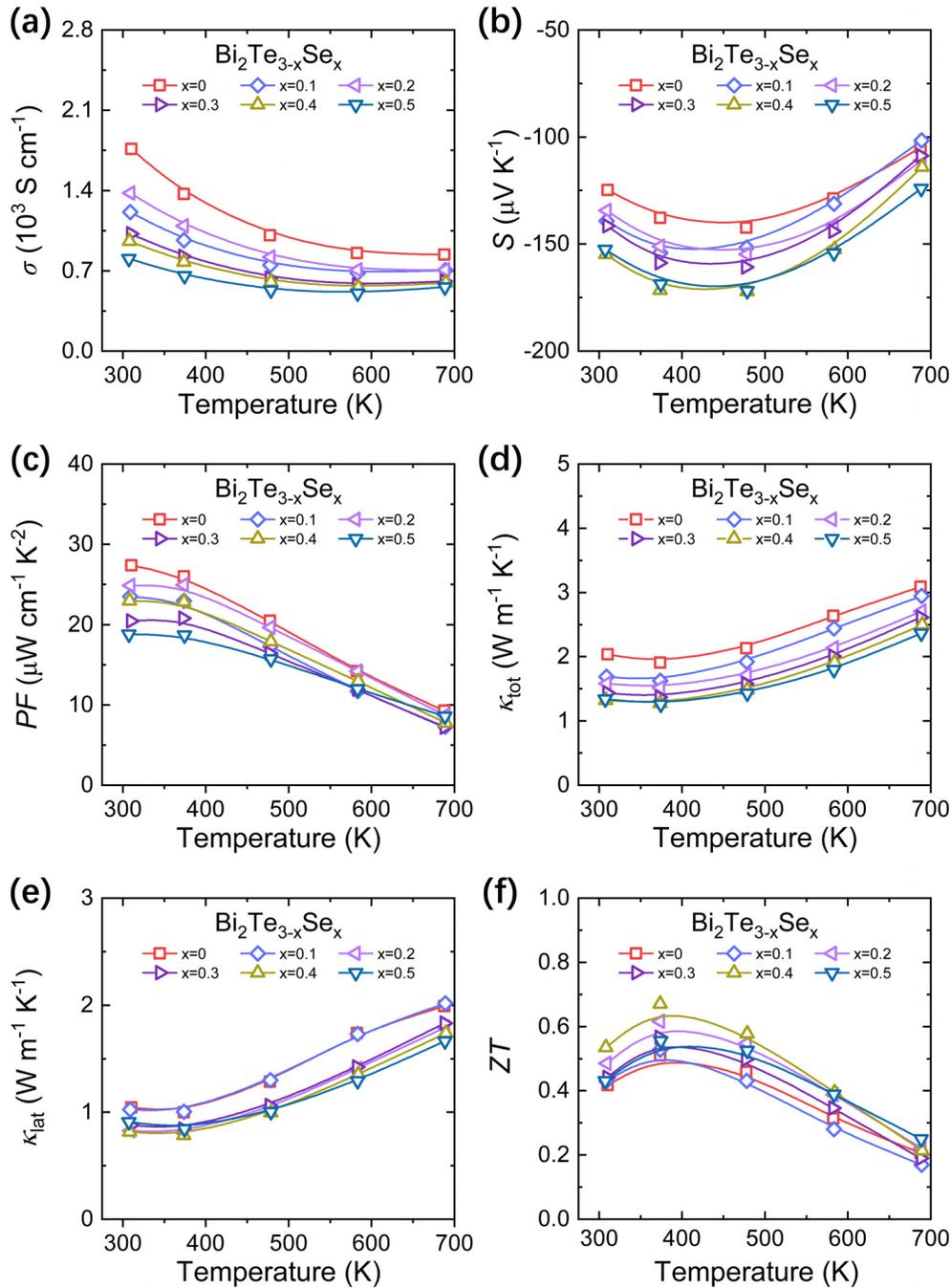

**FIG. 2.** Temperature-dependent thermoelectric transport properties of $Bi_2Te_{3-x}Se_x$ ($x = 0$–$0.5$). (a) Electrical conductivity. (b) Seebeck coefficient. (c) Power factor. (d) Total thermal conductivity. (e) Lattice thermal conductivity. (f) ZT values.

distribution, and good reproducibility. Remarkably, the annealed $Bi_2Te_{2.6}Se_{0.4}$ sample exhibits significantly enhanced thermoelectric performance compared to other n-type BTS compositions. As shown in Fig. 1(a), the sample achieves a high room-temperature carrier mobility ($\mu$) of 295.97 cm$^2$ V$^{-1}$ s$^{-1}$ and a large reciprocal $\kappa_{lat}$ of 2.19 W$^{-1}$ m K, outperforming reported n-type $Bi_2(Te,Se)_3$ compositions such as $Bi_2Te_{2.85}Se_{0.15}$,[15] $Bi_2Te_{2.7}Se_{0.3}$,[12,16–18] $Bi_2Te_{2.5}Se_{0.5}$,[19] $Bi_2Te_{2.4}Se_{0.6}$,[20,21] $Bi_2Te_{2.3}Se_{0.7}$,[14] and $Bi_2Te_{2.1}Se_{0.9}$.[22,23] These improvements are attributed to the synergistic effect of the MT-HP method and the subsequent annealing treatment (MT-HP-AN), as shown in

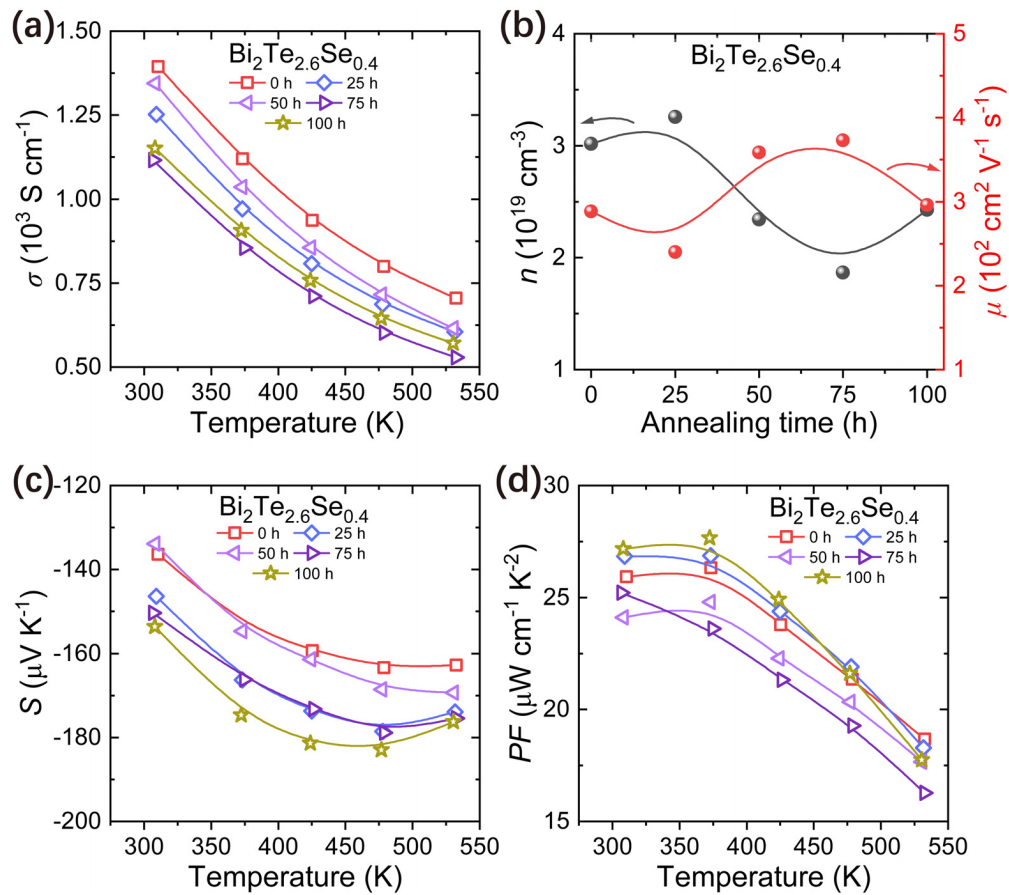

**FIG. 3.** Temperature-dependent electric transport properties of $Bi_2Te_{2.6}Se_{0.4}$ annealed at 723 K as a function of annealing time. (a) Electrical conductivity. (b) Carrier density and mobility. (c) Seebeck coefficient. (d) Power factor.

Fig. 1(b). Notably, whether prepared via MT-HP or MT-SPS, the annealed samples consistently exhibited enhanced thermoelectric performance compared to their unannealed counterparts, confirming the efficacy of annealing in optimizing the transport properties of n-type BTS. The optimized $Bi_2Te_{2.6}Se_{0.4}$ composition exhibits a high $ZT$ value of 1.06 at 373 K and a room-temperature $ZT_{RT}$ value of 0.78, resulting in an average $ZT_{ave}$ value of 0.99 at 300–423 K, among the highest for n-type $Bi_2(Te,Se)_3$ alloys [Figs. 1(c) and 1(d)]. The enhanced thermoelectric properties are attributed to high carrier mobility and reduced lattice thermal conductivity. Combined with a scalable MT-HP synthesis and annealing strategy, this work offers a practical route to high-performance n-type $Bi_2Te_3$-based thermoelectrics.

In this work, we first prepared a series of $Bi_2Te_{3-x}Se_x$ samples by MT-HP method with a hot pressing temperature of 673 K. PXRD patterns [Fig. S1(a)] confirm phase purity with all samples crystallizing in the rhombohedral $Bi_2Te_3$ structure. The gradual peak shift [Fig. S1(b)] with increasing Se content indicates lattice contraction due to substitution of Te with smaller Se atoms, without structural transformation. Thermoelectric measurements along both in-plane ($\perp$) and out-of-plane directions ($\parallel$) (Fig. S2) reveal superior performance in the in-plane direction. As shown in Fig. 2, increasing Se content decreases electrical conductivity while enhancing the absolute Seebeck coefficient ($|S|$) from 124.74 to 154.84 $\mu V\ K^{-1}$ at room temperature [Figs. 2(a) and 2(b)]. This behavior is attributed to the suppression of "donor-like" antisite $Bi'_{Te(Se)}$ defects.[24] However, due to the concurrent drop in conductivity, the power factor shows no significant improvement across the series [Fig. 2(c)].

$\kappa_{tot}$ and $\kappa_{lat}$ of $Bi_2Te_{3-x}Se_x$ are shown in Figs. 2(d) and 2(e). Se substitution effectively reduces $\kappa_{tot}$ by suppressing both electronic thermal conductivity $\kappa_{ele}$ and $\kappa_{lat}$ contributions. The decrease in $\kappa_{ele}$ correlates with the reduced electrical conductivity (Fig. S3). Moreover, the bandgap $E_g$, estimated using $E_g = 2e|S_{max}|T$,[25] increases with Se content due to the enhanced $|S_{max}|$, thereby suppressing bipolar diffusion. The reduction in $\kappa_{lat}$ is attributed to both suppressed bipolar thermal conduction and increased phonon scattering arising from Se-induced lattice distortion. As shown in Fig. 2(f), $Bi_2Te_{2.6}Se_{0.4}$ achieves a peak $ZT$ of 0.67 at 373 K, surpassing the conventionally favored $Bi_2Te_{2.7}Se_{0.3}$ composition.

Additionally, we also investigated the effect of hot pressing temperature on the thermoelectric performance of $Bi_2Te_{2.6}Se_{0.4}$ under a constant pressure of 50 MPa (Figs. S4 and S5). As shown in Fig. S4, the sample hot-pressed at 723 K exhibits the optimal thermoelectric performance. The HP process can be considered as an annealing treatment and we will discuss this amplified effect in detail in the later annealing process.

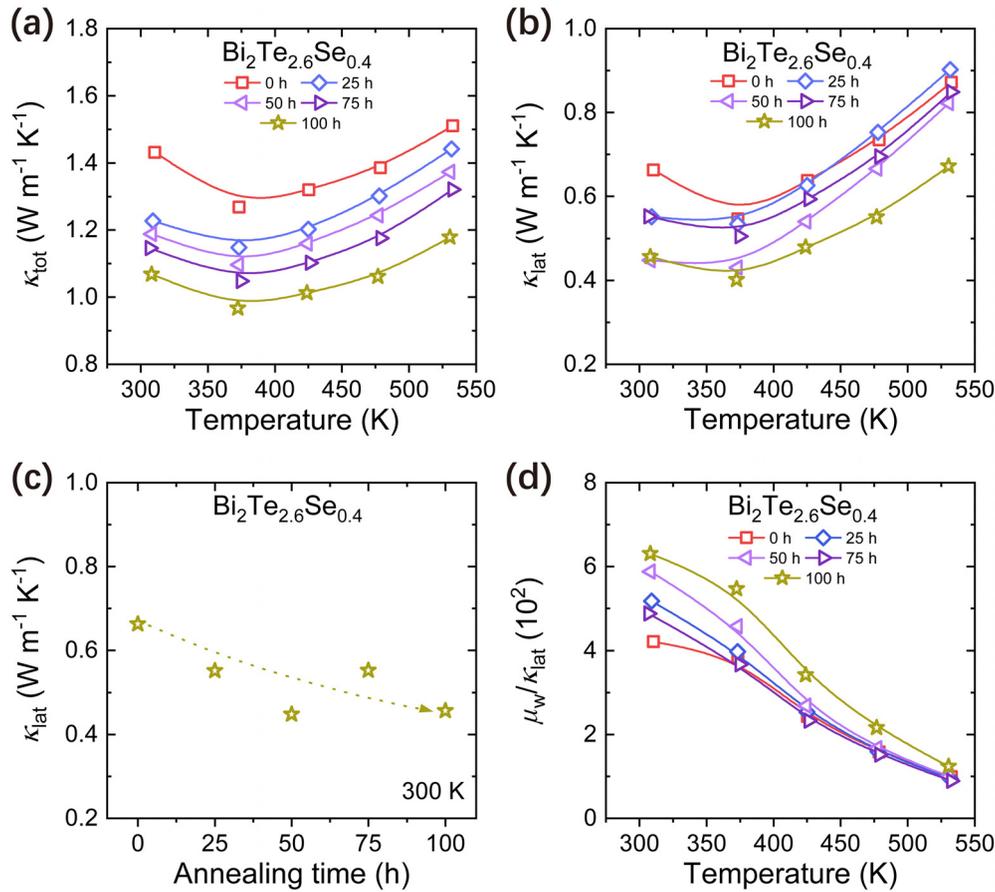

**FIG. 4.** Temperature-dependent thermal transport properties of $Bi_2Te_{2.6}Se_{0.4}$ annealed at 723 K as a function of annealing time. (a) Total thermal conductivity. (b) Lattice thermal conductivity. (c) Room-temperature lattice thermal conductivity as a function of annealing time for $Bi_2Te_{2.6}Se_{0.4}$. (d) The ratio of weighted carrier mobility to lattice thermal conductivity (the unit of $\mu_W/\kappa_{lat}$ is $cm^2\,V^{-1}\,s^{-1}/W\,m^{-1}\,K^{-1}$).

Figure 3 shows the electrical transport properties as a function of measuring temperature for the $Bi_2Te_{2.6}Se_{0.4}$ annealed at 723 K. As shown in Fig. 3(a), the electrical conductivity of all annealed samples is reduced compared to unannealed samples due to the reduced carrier density during annealing process in Fig. 3(b). As shown in Fig. 3(b), the carrier density and mobility exhibit opposite trends as the annealing time is prolonged. The annealed samples exhibited a reduction in carrier density from $3.02 \times 10^{19}$ to $2.43 \times 10^{19}\,cm^{-3}$, accompanied by an increase in carrier mobility from 288.44 to 295.97 $cm^2\,V^{-1}\,s^{-1}$ after being annealed for 100 h. The main reasons that caused the decrease of carrier density are the enhanced formation of antisite defect $Bi'_{Te(Se)}$ and cation vacancy $V'''_{Bi}$ caused by the volatilization of Te(Se) during the annealing process.[26] With the volatilization of Te(Se) during the annealing process, the evaporation of each Te(Se) leaves one Te(Se) vacancy $V^{\bullet\bullet}_{Te(Se)}$.[11] Subsequently, the antisite defect $Bi'\,Te(Se)$ and Bi vacancy $V'''\,Bi$ are produced as Bi atoms enter Te vacancies.[26] On the one hand, the "donor-like" defects[27] formed in the grinding and pressing process can be reduced by annealing treatment as it promotes grain rearrangement and recrystallization.[26,28,29]

As discussed above, with the extension of annealing time, the carrier density decreases. The decrease in carrier density is therefore is attributed to compensated electrons by holes and reduced "donor-like" defects, a finding consistent with other reports.[26,28] As shown in Fig. 3(c), the |S| values of all annealed samples are increased due to the decrease in carrier density. In addition, the variations of density of states (DOS) effective mass in Fig. S7 are associated with the volatilization of elements during the annealing process. Consequently, due to its enhanced |S| values and high electrical conductivity, a room-temperature PF value of 27.65 $\mu W\,cm^{-1}\,K^{-2}$ and a peak PF value of 27.17 $\mu W\,cm^{-1}\,K^{-2}$ at 373 K are obtained in the $Bi_2Te_{2.6}Se_{0.4}$ sample annealed for 100 h at 723 K, as shown in Fig. 3(d).

Figure 4 shows the temperature-dependent thermal transport properties of $Bi_2Te_{2.6}Se_{0.4}$ annealed at 723 K with different annealing time. As shown in Fig. 4(a), compared to the unannealed sample, the $\kappa_{tot}$ values of all annealed samples are reduced owing to the reduced carrier density and intensive point defects scattering introduced by enhanced cation vacancies during the annealing process. The $\kappa_{ele}$ values of annealed samples are lower than that of the unannealed sample, as shown in Fig. S8. This reduction is attributed to decreased electrical conductivity caused by a lower carrier density after annealing. Additionally, enhanced point defect scattering caused by element volatilization greatly hinders phonon propagation, resulting in a decrease in $\kappa_{lat}$ [Fig. 4(b)]. Notably, $\kappa_{lat}$ at high temperatures deviates from the conventional Umklapp scattering behavior ($\kappa_{lat} \propto T^{-1}$). Here, $\kappa_{lat}$ was derived by subtracting the electronic contribution from the total thermal conductivity. At elevated temperatures, bipolar thermal conduction becomes significant in $Bi_2(Te,Se)_3$-based materials due to thermally activated minority carriers. Since this bipolar contribution cannot be fully removed by the Wiedemann–Franz law estimation of $\kappa_{ele}$, it is included in the residual term, which gives rise to the apparent upturn in the derived $\kappa_{lat}$. The room-temperature lattice thermal conductivity in $Bi_2Te_{2.6}Se_{0.4}$ decreases with prolonged annealing time in Fig. 4(c). An exceptionally low value of 0.48 W m$^{-1}$ K$^{-1}$ can be achieved at room temperature in the $Bi_2Te_{2.6}Se_{0.4}$ sample annealed for 100 h. Here, the temperature-dependent ratio of weighted carrier mobility to lattice thermal conductivity ($\mu_W/\kappa_{lat}$) is calculated out to evaluate the contribution of annealing to thermoelectrical transport properties. The parameter $\mu_W$ can be calculated with measured electrical conductivity and Seebeck coefficient using the following relationships:[30]

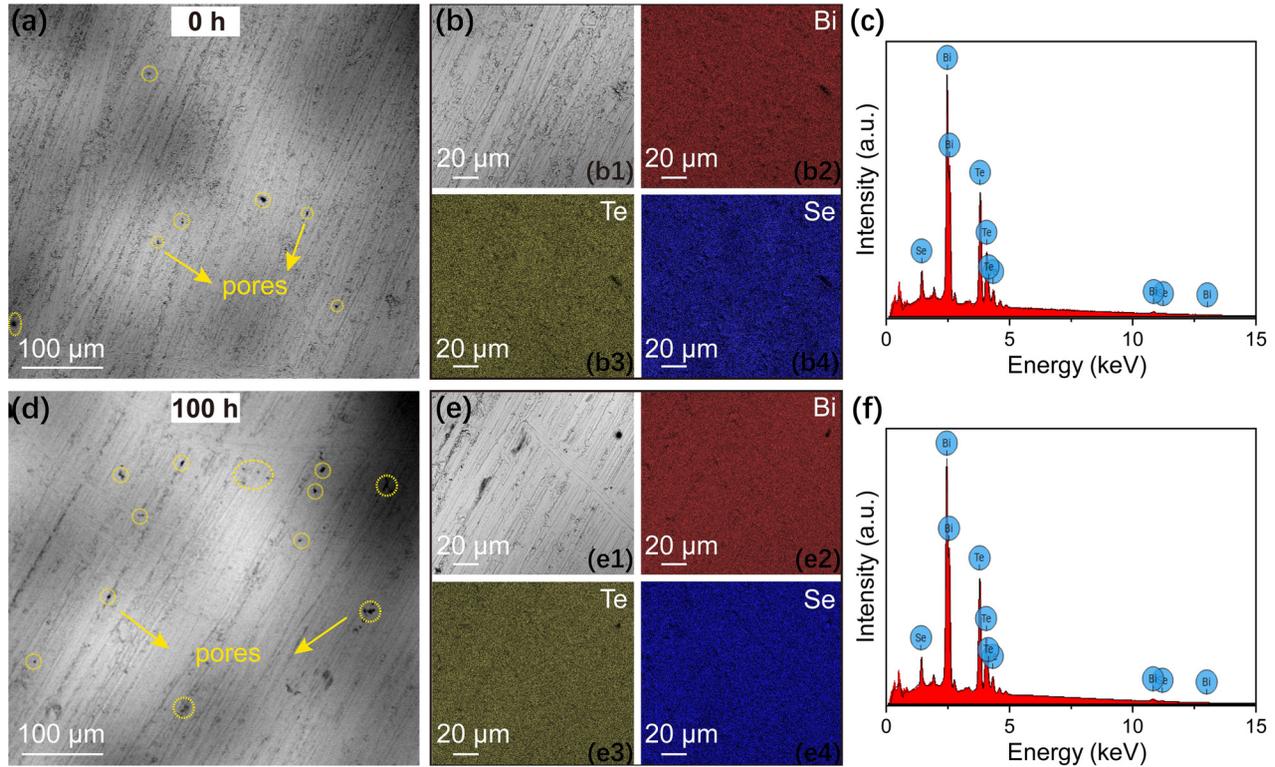

FIG. 5. SEM graphs, EDS mapping, and EDS spectra of $Bi_2Te_{2.6}Se_{0.4}$ bulks before and after annealing. (a)–(c) Without annealing. (d)–(f) Annealing at 723 K for 100 h.

TABLE I. Chemical composition of $Bi_2Te_{2.6}Se_{0.4}$ bulks before and after annealing.

| Annealing time (h) | Bi/Te/Se ratio | Bi/(Te+Se) ratio |
|---|---|---|
| 0 | 2/2.61/0.31 | 2/2.92 |
| 100 | 2/2.35/0.27 | 2/2.62 |

$$\mu_W = \frac{3\sigma}{8\pi e F_0(\eta)} \left(\frac{h^2}{2m_e k_B T}\right)^{3/2}, \quad (1)$$

$$F_n(\eta) = \int_0^\infty \frac{x^n}{1+e^{x-\eta}} dx, \quad (2)$$

$$S = \pm \frac{k_B}{e} \left(\frac{(r+5/2)F_{r+3/2}(\eta)}{(r+3/2)F_{r+1/2}(\eta)} - \eta\right), \quad (3)$$

where $e$, $h$, $m_e$, and $k_B$ denote the unit charge, Planck constant, electron mass, and the Boltzmann constant, respectively. $F_n(\eta)$ is the Fermi integral, $\eta$ represents the reduced Fermi level, and $r$ is the scattering factor which equals $-1/2$ when the acoustic scattering mechanism dominates. After annealing, the $\mu_W/\kappa_{lat}$ value of all samples is significantly enhanced, as shown in Fig. 4(d). This enhanced $\mu_W/\kappa_{lat}$ value indicates that the atomic defects introduced into the matrix during the annealing process can intensify phonon scattering while maintaining high carrier mobility, thereby realizing synergistic optimization between carrier and phonon transport properties.

To investigate the effects of annealing on microstructure and composition, SEM and EDS mapping were performed on $Bi_2Te_{2.6}Se_{0.4}$ samples before and after 100 h annealing. As shown in Fig. 5, micrographs perpendicular to the pressing direction reveal that pore density and size increase after annealing at 723 K [Figs. 5(a) and 5(b1) vs 5(d) and 5(e1)], consistent with density measurements (Table S3). The reduced density likely originates from volatilization of Te(Se) during annealing, in agreement with prior reports.[26]

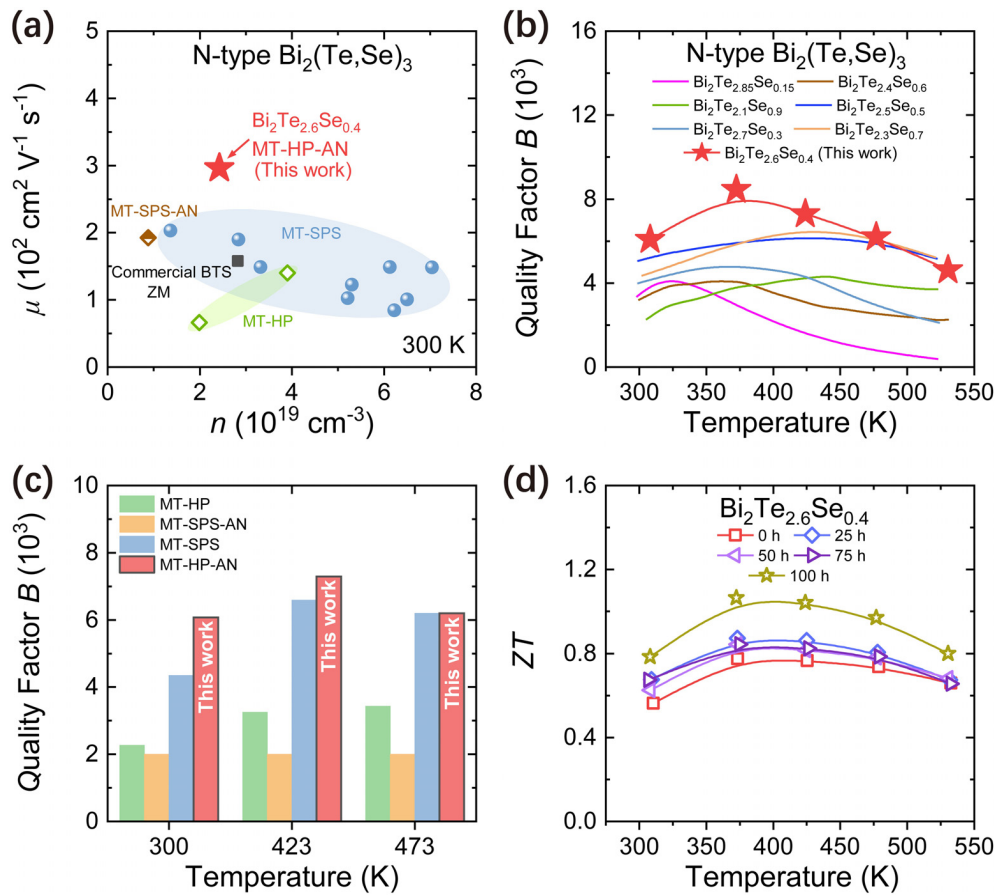

**FIG. 6.** Comparison of thermoelectric transport properties in n-type $Bi_2(Te,Se)_3$-based thermoelectric materials. (a) Carrier mobility as a function of carrier density at room temperature with different manufacturing process, including zone melting (ZM), MT-HP, MT-SPS, MT-SPS-AN. Data from Refs. 12, 14–23 and 31. (b) and (c) Quality factor $B$ (the unit of $B$ is $cm^2\,V^{-1}\,s^{-1}/W\,m^{-1}\,K^{-1}$). Data from Refs. 14, 15, 17, 19, and 21–23. (d) $ZT$ values of $Bi_2Te_{2.6}Se_{0.4}$ with different annealing time.

EDS elemental mapping [Figs. 5(b) and 5(c) and 5(e) and 5(f)] shows clear evidence of element depletion in the pore regions of annealed samples, correlating with the reduction in $\kappa_{lat}$. Notably, both unannealed and annealed samples exhibit deviation from the nominal $Bi_2Te_{2.6}Se_{0.4}$ composition (Table I), likely due to Se loss during synthesis steps such as flame sealing, grinding, and hot pressing, attributable to the lower evaporation enthalpy of Se (37.70 kJ mol$^{-1}$) compared to Te (52.55 kJ mol$^{-1}$).[11] After 100 h annealing, the Bi/(Te+Se) ratio further increases, indicating enhanced volatilization of Te(Se) during prolonged heat treatment.

As mentioned above, appropriate annealing treatment can optimize the thermoelectric performance of n-type BTS. As shown in Fig. 6(a), the $Bi_2Te_{2.6}Se_{0.4}$ sample prepared by MT-HP-AN has a suitable room-temperature carrier density of $2.43 \times 10^{19}\,cm^{-3}$ and a high carrier mobility of 295.97 $cm^2\,V^{-1}\,s^{-1}$. Such high carrier mobility is superior to that of commercial BTS prepared by zone melting (ZM)[31] and other BTS samples prepared by MT-HP,[12,22] MT-SPS,[14–20,23] and MT-SPS-AN.[21] Since appropriate annealing process can synergistically optimize thermal and electrical transport, we plotted the temperature-dependent quality factor $B$ to better evaluate this contribution, which is defined by the following relation:[32]

$$B = 9\frac{\mu_W}{\kappa_{lat}}\left(\frac{T}{300}\right)^{5/2}. \quad (4)$$

As shown in Fig. 6(b), it can be seen that the $B$ values of the $Bi_2Te_{2.6}Se_{0.4}$ sample here surpass those of other n-type $Bi_2(Te,Se)_3$ samples with different Se content, including $Bi_2Te_{2.85}Se_{0.15}$,[15] $Bi_2Te_{2.7}Se_{0.3}$,[17] $Bi_2Te_{2.5}Se_{0.5}$,[19] $Bi_2Te_{2.4}Se_{0.6}$,[21] $Bi_2Te_{2.3}Se_{0.7}$,[14] and $Bi_2Te_{2.1}Se_{0.9}$.[23] This phenomenon results from different manufacturing processes (MT-HP,[22] MT-SPS,[14] MT-SPS-AN[21]) and is demonstrated in Fig. 6(c), indicating that annealing treatment in BTS alloys is an effective method to simultaneously boost electrical properties and weaken thermal transport properties. As a result, after annealing treatment, the final $ZT$ value in n-type BTS undergoes significant enhancement in the whole working temperature range. A room-temperature $ZT_{RT}$ value of 0.78 and a maximum $ZT$ value of 1.06 at 373 K have been realized in the $Bi_2Te_{2.6}Se_{0.4}$ sample annealed for 100 h, which results in a high average $ZT_{ave}$ value of 1.0 at 300–423 K, shown in Fig. 6(d). The introduction of atomic defects in $Bi_2Te_{2.6}Se_{0.4}$ during the annealing process can lead to a reduction of lattice thermal conductivity while simultaneously maintaining a relatively high power factor, which greatly optimizes the thermoelectric transport properties. The present superior thermoelectric properties at low to medium temperature indicate that optimized $Bi_2Te_{2.6}Se_{0.4}$ has great potential for thermoelectric cooling applications in advanced information systems.

In summary, the thermoelectric performance of Se-alloyed n-type $Bi_2Te_3$-based (BTS) polycrystals was found to be significantly enhanced via a simple annealing treatment that effectively decouples carrier and phonon transport. The optimal composition of $Bi_2Te_{2.6}Se_{0.4}$ was identified to suppress $\kappa_{lat}$ through defect engineering. Systematic annealing at 723 K tuned carrier density and mobility through the cation vacancy ($V'''_{Bi}$) formation induced by Te(Se) volatilization. After 100 h of annealing, the sample exhibited reduced carrier density, enhanced mobility (295.97 $cm^2 V^{-1} s^{-1}$), and improved power factor. Concurrently, increased point defects strengthened phonon scattering, leading to significantly reduced $\kappa_{lat}$. These combined effects yielded a peak $ZT$ of 1.06 and an average $ZT_{ave}$ of 0.99 at 300–423 K. This study presents a practical post-synthesis route for optimizing thermoelectric performance in n-type BTS alloys, demonstrating strong potential for chip-level thermal management and cooling applications.

See the supplementary material for detailed experimental procedures, additional characterization data, and description of the theoretical calculation methods.

We acknowledge receiving financial support from the National Natural Science Foundation of China (Grant Nos. 52250191 and 52172236). This work was also supported by the Sichuan Provincial Natural Science Foundation (Grant No. 2025ZNSFSC0385) and by the Key Research and Development Program of the Ministry of Science and Technology under Grant No. 2023YFB4604100. We also acknowledge receiving support from the HPC Platform, Xi'an Jiaotong University.

## AUTHOR DECLARATIONS
### Conflict of Interest

The authors have no conflicts to disclose.

### Author Contributions

Xinxiu Cheng and Liqing Xu contributed equally to this work.

**Xinxiu Cheng:** Data curation (equal); Writing – original draft (equal). **Liqing Xu:** Data curation (equal); Writing – original draft (equal). **Zhibin Gao:** Conceptualization (equal); Formal analysis (equal); Funding acquisition (equal); Project administration (equal); Supervision (equal); Writing – review & editing (equal). **Wei Liu:** Formal analysis (supporting). **Zhanxiang Yin:** Data curation (supporting). **XiangDong Ding:** Project administration (equal). **Yu Xiao:** Conceptualization (equal); Formal analysis (equal); Funding acquisition (equal); Project administration (equal); Supervision (equal); Writing – review & editing (equal).

## DATA AVAILABILITY

The data that support the findings of this study are available from the corresponding authors upon reasonable request.

### REFERENCES

[1]D. W. Yang, Y. B. Xing, J. Wang, K. Hu, Y. N. Xiao, K. C. Tang, J. N. Lyu, J. H. Li, Y. T. Liu, P. Zhou, Y. Yu, Y. G. Yan, and X. F. Tang, Interdiscip. Mater. **3**(2), 326 (2024); S. P. Zhan, S. L. Bai, Y. T. Qiu, L. Zheng, S. N. Wang, Y. C. Zhu, Q. Tan, and L. D. Zhao, Adv. Mater. **36**(47), 2412967 (2024); J. Mao, G. Chen, and Z. F. Ren, Nat. Mater. **20**(4), 454 (2021); L. Q. Xu, Y. Xiao, S. N. Wang, B. Cui, D. Wu, X. D. Ding, and L. D. Zhao, Nat. Commun. **13**(1), 6449 (2022); J. Mao, H. T. Zhu, Z. W. Ding, Z. H. Liu, G. A. Gamage, G. Chen, and Z. F. Ren, Science **365**(6452), 495 (2019); B. Qin and L. D. Zhao, Science **378**(6622), 832 (2022).

[2]I. T. Witting, T. C. Chasapis, F. Ricci, M. Peters, N. A. Heinz, G. Hautier, and G. J. Snyder, Adv. Electron. Mater. **5**(6), 1800904 (2019).

[3]Y. X. Qin, B. C. Qin, D. Y. Wang, C. Chang, and L. D. Zhao, Energy Environ. Sci. **15**(11), 4527 (2022).

[4]Y. Xiao, Mater. Lab **1**(3), 220025 (2022); Z. H. Liu, W. H. Gao, F. K. Guo, W. Cai, Q. Zhang, and J. H. Sui, Mater. Lab **1**(2), 220003 (2022); B. C. Qin and L. D. Zhao, Mater. Lab. **1**(1), 220004 (2022); Z. T. Liu, T. Hong, L. Q. Xu, S. N. Wang, X. Gao, C. Chang, X. D. Ding, Y. Xiao, and L. D. Zhao, Interdiscip. Mater. **2**(1), 161 (2023); H. Y. Xie, L. D. Zhao, and M. G. Kanatzidis, Interdiscip. Mater. **3**(1), 5 (2024).

[5]Y. L. Huang, T. Lyu, M. T. Zeng, M. R. Wang, Y. Yu, C. H. Zhang, F. S. Liu, M. Hong, and L. P. Hu, Interdiscip. Mater. **3**(4), 607 (2024).

[6]Y. Xiao and L. D. Zhao, Science **367**(6483), 1196 (2020).

[7]L. J. Xie, L. Yin, Y. Yu, G. Y. Peng, S. W. Song, P. J. Ying, S. T. Cai, Y. X. Sun, W. J. Shi, H. Wu, N. Qu, F. K. Guo, W. Cai, H. J. Wu, Q. Zhang, K. Nielsch, Z. F. Ren, Z. H. Liu, and J. H. Sui, Science **382**(6673), 921 (2023).

[8]B. Xu, M. T. Agne, T. L. Feng, T. C. Chasapis, X. L. Ruan, Y. L. Zhou, H. M. Zheng, J. H. Bahk, M. G. Kanatzidis, G. J. Snyder, and Y. Wu, Adv. Mater. **29**(10), 1605140 (2017); X. Y. Liu, D. Y. Wang, H. J. Wu, J. F. Wang, Y. Zhang, G. T. Wang, S. J. Pennycook, and L. D. Zhao, Adv. Funct. Mater. **29**(3), 1806558 (2019); B. Poudel, Q. Hao, Y. Ma, Y. C. Lan, A. Minnich, B. Yu, X. Yan, D. Wang, A. Muto, D. Vashaee, X. Chen, J. Liu, M. S. Dresselhaus, G. Chen, and Z. Ren, Science **320**(5876), 634 (2008).

[9]S. I. Kim, K. H. Lee, H. A. Mun, H. S. Kim, S. W. Hwang, J. W. Roh, D. J. Yang, W. H. Shin, X. S. Li, Y. H. Lee, G. J. Snyder, and S. W. Kim, Science **348**(6230), 109 (2015); Y. Pan, Y. Qiu, I. Witting, L. Zhang, C. G. Fu, J. W. Li, Y. Huang, F. H. Sun, J. Q. He, G. J. Snyder, C. Felser, and J. F. Li, Energy Environ. Sci. **12**(2), 624 (2019); Y. Wu, Y. Yu, Q. Zhang, T. J. Zhu, R. S. Zhai, and X. B. Zhao, Adv. Sci. **6**(21), 1901702 (2019).

[10]T. Chen, H. W. Ming, X. Y. Qin, C. Zhu, Y. Chen, L. Ai, D. Li, Y. S. Zhang, H. X. Xin, and J. Zhang, Inorg. Chem. Front. **9**(20), 5386 (2022); D. R. Liu, S. L. Bai, Y. Wen, J. Y. Peng, S. B. Liu, H. N. Shi, Y. C. Li, T. Hong, H. Q. Liang, Y. X. Qin, L. Z. Su, X. Qian, D. Y. Wang, X. Gao, Z. H. Ding, Q. Cao, Q. Tan, B. C. Qin, and L. D. Zhao, Natl. Sci. Rev. **12**(2), nwae448 (2024); I. T. Witting, F. Ricci, T. C. Chasapis, G. Hautier, and G. J. Snyder, Research **2020**, 4361703.

[11]W. S. Liu, Q. Y. Zhang, Y. C. Lan, S. Chen, X. Yan, Q. Zhang, H. Wang, D. Z. Wang, G. Chen, and Z. F. Ren, Adv. Energy Mater **1**(4), 577 (2011).

[12]X. Y. Chen, J. Li, Q. Shi, Y. Y. Chen, H. J. Gong, Y. P. Huang, L. Lin, D. Ren, B. Liu, and R. Ang, ACS Appl. Mater. Interfaces **13**(49), 58781 (2021).

[13]A. Haruna, Y. Luo, W. Li, M. An, P. Fu, X. Li, Q. Jiang, and J. Yang, J. Mater. Chem. A **12**(7), 4221 (2024); S. Chen, K. Cai, F. Li, and S. Shen, J. Electron. Mater. **43**(6), 1966 (2014).

[14]B. Q. Jia, H. Zhang, F. D. Zhang, H. S. Li, B. P. Ma, W. S. Wang, Y. L. Shi, X. L. Chao, Z. P. Yang, and D. Wu, ACS Appl. Mater. Interfaces **16**(44), 60588 (2024); C. H. Zhang, C. X. Zhang, H. K. Ng, and Q. Q. Xiong, Sci. China Mater. **62**(3), 389 (2019); B. Zhu, X. X. Liu, Q. Wang, Y. Qiu, Z. Shu, Z. T. Guo, Y.

Tong, J. Cui, M. Gu, and J. Q. He, Energy Environ. Sci. **13**(7), 2106 (2020); M. Y. Wang, Z. L. Tang, T. J. Zhu, and X. B. Zhao, RSC Adv. **6**(101), 98646 (2016); L. Y. Miao, X. Lu, Q. Zhang, X. J. Tan, L. D. Chen, K. K. Pang, R. Y. Li, Q. Q. Sun, M. Wang, P. Sun, J. H. Wu, G. Q. Liu, Z. L. Song, and J. Jiang, J. Mater. Sci. Technol. **223**, 114 (2025).

[15]S. T. Han, P. Rimal, C. H. Lee, H. S. Kim, Y. Sohn, and S. J. Hong, Intermetallics **78**, 42 (2016).

[16]R. W. Yin, H. F. Zhou, Y. Wang, L. Pan, C. C. Chen, S. P. He, and C. L. Wan, J. Alloys Compd. **1011**, 178397 (2025).

[17]S. G. Lin, J. Li, H. Yan, X. F. Meng, Q. P. Xiang, H. Jing, X. X. Chen, and C. T. Yang, Materials **17**(8), 1919 (2024).

[18]A. Q. Zhao, H. Liu, T. Sun, Y. D. Lang, C. C. Chen, L. Pan, and Y. F. Wang, J. Alloys Compd. **982**, 173806 (2024); L. Li, P. Wei, M. J. Yang, W. T. Zhu, X. L. Nie, W. Y. Zhao, and Q. J. Zhang, Sci. China Mater. **66**(9), 3651 (2023); Y. J. Jung, H. S. Kim, J. H. Won, M. Kim, M. Kang, E. Y. Jang, N. V. Binh, S. I. Kim, K. S. Moon, J. W. Roh, W. H. Nam, S. M. Koo, J. M. Oh, J. Y. Cho, and W. H. Shin, Materials **15**(6), 2284 (2022).

[19]Y. Xiao, J. Y. Yang, Q. H. Jiang, L. W. Fu, Y. B. Luo, D. Zhang, and Z. W. Zhou, J. Mater. Chem. A **3**(44), 22332 (2015).

[20]Z. Y. Huang, H. Q. Li, X. Y. Wang, W. Jiang, and F. Q. Zu, J. Mater. Sci. **30**(16), 15018 (2019).

[21]F. Hao, T. Xing, P. F. Qiu, P. Hu, T. Wei, D. R. Ren, X. D. Shi, and L. D. Chen, ACS Appl. Mater. Interfaces **10**(25), 21372 (2018).

[22]H. Cho, J. H. Yun, J. H. Kim, S. Y. Back, H. S. Lee, S. J. Kim, S. Byeon, H. Jin, and J. S. Rhyee, ACS Appl. Mater. Interfaces **12**(1), 925 (2020).

[23]D. Li, X. Y. Qin, J. Zhang, C. J. Song, Y. F. Liu, L. Wang, H. X. Xin, and Z. M. Wang, RSC Adv. **5**(54), 43717 (2015).

[24]J. Jiang, Y. L. Li, G. J. Xu, P. Cui, T. Wu, L. D. Chen, and G. Wang, Acta Phys. Sin. **56**(5), 2858 (2007).

[25]H. J. Goldsmid and J. W. Sharp, J. Electron. Mater. **28**(7), 869 (1999).

[26]L. D. Zhao, B. P. Zhang, W. S. Liu, H. L. Zhang, and J. F. Li, J. Alloys Compd. **467**(1–2), 91 (2009).

[27]R. Y. Li, Q. Y. Pan, Q. Zhang, M. Wang, K. K. Pang, L. Y. Miao, X. J. Tan, H. Y. Hu, J. H. Wu, G. Q. Liu, and J. Jiang, Small **21**(3), 2408794 (2025).

[28]L. D. Zhao, B. P. Zhang, J. F. Li, M. Zhou, and W. S. Liu, Physica B **400**(1–2), 11 (2007).

[29]D. M. Lee, C. H. Lim, D. C. Cho, Y. S. Lee, and C. H. Lee, J. Electron. Mater. **35**(2), 360 (2006); J. M. Schultz, J. P. McHugh, and W. A. Tiller, J. Appl. Phys. **33**(8), 2443 (1962).

[30]G. J. Snyder, A. H. Snyder, M. Wood, R. Gurunathan, B. H. Snyder, and C. N. Niu, Adv. Mater. **32**(25), 2001537 (2020).

[31]J. Y. Peng, D. R. Liu, S. L. Bai, Y. Wen, H. Q. Liang, L. Z. Su, X. Qian, D. Y. Wang, X. Gao, Z. H. Ding, Q. Cao, Y. L. Pei, B. C. Qin, and L. D. Zhao, Adv. Energy Mater. **15**(18), 2404653 (2025).

[32]M. Ohta, D. Y. Chung, M. Kunii, and M. G. Kanatzidis, J. Mater. Chem. A **2**(47), 20048 (2014).